\documentclass[pra,superscriptaddress,twocolumn,nofootinbib,longbibliogrpahy]{revtex4-2}

\usepackage{enumerate}
\usepackage{amsfonts,amssymb,amsmath}
\usepackage[]{graphics,graphicx,epsfig}
\usepackage{amsthm}
\usepackage{amssymb}
\usepackage{mathbbol}
\usepackage{graphicx}
\usepackage{dcolumn}
\usepackage{natbib}
\usepackage{color}
\usepackage{multirow}
\usepackage{ulem}
\usepackage{bm}
\usepackage{url}

\usepackage[colorlinks = true, citecolor=blue, linkcolor=blue, urlcolor=blue]{hyperref}
\newcommand*{\fullref}[1]{\hyperref[{#1}]{\autoref*{#1} \nameref*{#1}}}

\begin{document}

\title{Qubit from the classical collision entropy}

\author{Kelvin Onggadinata}
\email{kelvin.onggadinata@u.nus.edu}
\address{Centre for Quantum Technologies,
National University of Singapore, 3 Science Drive 2, 117543 Singapore,
Singapore}
\address{Department of Physics,
National University of Singapore, 3 Science Drive 2, 117543 Singapore,
Singapore}

\author{Pawe{\l} Kurzy{\'n}ski}
\address{Institute of Spintronics and Quantum Information, Faculty of Physics, Adam Mickiewicz University, Uniwersytetu Pozna{\'n}skiego 2, 61-614 Pozna\'n, Poland}
\address{Centre for Quantum Technologies,
National University of Singapore, 3 Science Drive 2, 117543 Singapore,
Singapore}

\author{Dagomir Kaszlikowski}
\email{phykd@nus.edu.sg}
\address{Centre for Quantum Technologies,
National University of Singapore, 3 Science Drive 2, 117543 Singapore,
Singapore}
\address{Department of Physics,
National University of Singapore, 3 Science Drive 2, 117543 Singapore,
Singapore}

\date{\today}

\begin{abstract}

An orthodox formulation of quantum mechanics relies on a set of postulates in Hilbert space supplemented with rules to connect it with classical mechanics such as quantisation techniques, correspondence principle, etc. Here we deduce a qubit and its dynamics straightforwardly from a discrete deterministic dynamics and conservation of the classical collision entropy. No Hilbert space is required although it can be inferred from this approach if necessary.

\end{abstract}

\maketitle

\section{Introduction}

Soon after its inception, physicists encoded quantum theory in complex Hilbert space together with a set of postulates to link it with experiment. Departure from classical physics was bridged with a set of techniques and prescriptions of how to `quantise' familiar classical objects such as Hamiltonian, Lagrangian, momentum, position, etc. Although this approach works `good enough', it lacks clear physical meaning unlike, for instance, general relativity theory, entirely derived from two experimentally falsifiable assumptions: relativity and equivalence principle. One's desire for more intuitive and simpler postulates to construct quantum theory is understandable in this context.

In the last few decades, some researchers found alternative axioms for quantum theory. Hardy \cite{hardy2001quantum} proposed five axioms which later expanded and framed as generalized probability theory (GPT) by Barrett \cite{barrett2007information}. All these alternative formulations of quanta are forced to use quasi-probability theory (also known as signed measures \cite{szekely2005half} or, colloquially, as negative probabilities) to account for observed randomness of microscopic phenomena \cite{ferrie2010necessity}. Note that GPT is constructed such that one never has to assign negative probabilities to measurement outcomes. This eliminates any ontological discussions about the meaning of negative probabilities, a practice we follow in this paper.     

Quasi-probabilities are almost as old as quanta itself thanks to Wigner who introduced them in 1932 \cite{wigner1932quantum}. Since then the idea has been discussed by many authors in diverse contexts \cite{szekely2005half,muckenheim1986review,dirac1942bakerian,feynman1987negative,bartlett1945negative,burgin2010interpretations}. The most recent developments in the field suggest that quasi-probabilities could be viewed as a fundamental resource in {\it quantum non-locality} \cite{abramsky2004operational, alsafi2013simulating, halliwell2013negative, oas2014exploring, onggadinata2021local} and {\it quantum computation} \cite{galvao2005discrete, veitch2012negative, veitch2014resource, howard2014contextuality, delfosse2015wigner, kaszlikowski2021little}, and used extensively in quantum optics \cite{glauber1963coherent, sudarshan1963equivalence, alonso2011wigner}.

Here we propose a simple information-theoretic postulate from which we can derive a discrete four dimensional quasi-stochastic system equivalent to a qubit and its dynamics. Curiously, with no further restrictions, our postulate allows universal-NOT operation but we can get rid of it if we assume continuity of reversible quasi-stochastic processes. One of the interesting aspects of this approach is that we can reconstruct discrete quantum system without invoking Hilbert space at all. The information-theoretic flavour of our proposal falls closely within the proximity of {\it information causality} and its subsequent generalisations \cite{pawlowski2009information, barnum2010entropy, alsafi2011information}.

\section{Deterministic dynamics and Renyi entropies}

Consider a particle on a one-dimensional lattice with $d$ vertices (enumerated by $i=1,2,\dots, d$) and periodic boundary conditions. Or, if you prefer, you can see it as an abstract dynamical system whose states are single-vertex occupations, i.e., bit strings like this $(000\dots 01000\dots 0)$, where $1$ denotes the $i$-th vertex occupation.  

The simplest possible discrete deterministic dynamics is when the particle starts at some vertex $i$ and in every step hops to the next vertex: $i\rightarrow i+1\rightarrow i+2\rightarrow \dots$. We represent the particle’s state as a $d$-dimensional basis vector $\mathbf{e}$ (only one entry equals to $1$ with others equal to $0$) and its dynamics as a permutation matrix $\Pi$. After $k$ steps, the initial state $\mathbf{e}$ evolves to $\mathbf{e}^{(k)}=\Pi^{k}\mathbf{e}$. 

If the initial conditions are uncertain, $\mathbf{e}$ becomes a probability vector $\mathbf{p}=[p_1,p_2,\dots,p_d]^T$ where $p_i$ ($p_i\geq 0$ and $\sum_{i=1}^dp_i=1$) is the probability of finding the particle in the vertex $i$. The dynamics stays the same, i.e., after $k$ steps the state changes to $\mathbf{p}^{(k)}=\Pi^{k}\mathbf{p}$. It is rather obvious to observe a simple information-theoretic property of such dynamics: 
$$\sum_{i=1}^dp_i^{\alpha}=\sum_{i=1}^d\left([\Pi^k\mathbf{p}]_i\right)^{\alpha}$$ for every $k$ and $\alpha\geq 0$, i.e., $\sum_{i=1}^dp_i^{\alpha}$ is constant for any initial probability distribution $\mathbf{p}$. More generality we can express this with Renyi-$\alpha$ entropies, $H_\alpha(\mathbf{p})$, as
\begin{equation}
    H_{\alpha}(\Pi^k\mathbf{p})=H_{\alpha}(\mathbf{p})\, ,
\end{equation} 
where $H_{\alpha}$ is 
\begin{equation}
    H_{\alpha}(\mathbf{p})=\frac{1}{1-\alpha}\log{\left(\sum_{i=1}^dp_i^{\alpha}\right)}\, .
\end{equation}
In fact, any permutation conserves all the Renyi entropies. Introducing Renyi entropies may look unnecessary but it does give us an option to study Shannon entropy for $\alpha=1$. It is physically obvious that Shannon entropy is constant for any initial state $\mathbf{p}$ and deterministic dynamics.

We now ask the fundamental question in this paper: {\it Can we extend deterministic dynamics in our model to some other dynamics, not necessarily deterministic, such that some Renyi entropies remain constant for any given initial state $\mathbf{p}$?}

A `yes’ answer would mean there is a family of dynamics $S$ and states such that (i) $S\mathbf{p}$ remains a proper positive probability distribution and, (ii) there is a range of $\alpha$ where
\begin{equation}\label{eq: renyi entropy conservation}
H_{\alpha}(S\mathbf{p})=H_{\alpha}(\mathbf{p})
\end{equation}
for any $\mathbf{p}$ for which $S\mathbf{p}$ is a proper positive probability distribution.

We can already expect certain features of this extension. First, any extended dynamics $S$ must be a $d\times d$ matrix whose rows sum up to one or else the probabilities $S\mathbf{p}$ would not be normalized. However, we cannot guarantee that $S$’s elements are all positive, making $S$ at least a quasi-stochastic matrix if not a quasi-bistochastic one.

To sum up, at this moment, we have a well defined, physically motivated mathematical problem. In the next section, we provide a solution of deep physical significance: we recover qubit and its dynamics without Hilbert space.

\section{Generalised dynamics and Renyi entropies}

The problem formulated in the previous section is difficult to solve. However, we found an important solution for the Renyi entropy with $\alpha=2$. This entropy is called {\it collision entropy} in the literature and it reads:
\begin{equation}
H_2(\mathbf{p})=-\log{|\mathbf{p}|^2}\, ,
\end{equation}
where $|\mathbf{p}| = \sqrt{\sum_{i=1}^d p_i^2}$, i.e., it is the geometric length of a $d$-dimensional vector $\mathbf{p}$. 

If you want to satisfy Eq. (\ref{eq: renyi entropy conservation}), i.e., keep the collision entropy invariant under a generalised dynamics $S$, you must have 
$$|S\mathbf{p}|^2=|\mathbf{p}|^2\, .$$
This is possible only if transposition of the matrix $S$ is its own inverse because of a trivial observation
$$|S\mathbf{p}|^2=(S\mathbf{p})\cdot(S\mathbf{p})=\mathbf{p}\cdot(S^TS\mathbf{p}) = \mathbf{p}\cdot\mathbf{p} = |\mathbf{p}|^2$$
$$\Rightarrow \quad S^T = S^{-1}$$
In other words, orthogonal matrices produce the new generalized dynamics.

Next, we need to find the minimal dimension $d$ where this is possible. It helps to note that if $S$ is a valid dynamics so is $S^T$ and thus $S$ must have columns and rows summing up to one if we want to keep $S\mathbf{p}$ a proper probability vector. This is only possible if some of the elements in rows and columns are negative because the off-diagonal terms in $SS^T$ must be equal to zero. Thus $S$ must be a quasi-bistochastic matrix.

For $d=2$ the most general quasi-bistochastic matrix reads
\begin{equation}
    S = \begin{bmatrix}
    q & 1-q \\ 1-q & q
    \end{bmatrix}\, ,
\end{equation}
where $q$ is an arbitrary real number. It is easy to see that orthogonality is only true for $q=0$ or $q=1$, which makes it a permutation. Hence, no new dynamics is observed for $d=2$.


The situation changes for $d=3$. We start with the generalised Birkhoff-von Neumann decomposition \cite{onggadinata2021local} of quasi-bistochastic matrices
\begin{equation}
S=\sum_{k=0}^2q_k\Pi^k+\sum_{k=0}^2r_k\Pi^kR\, ,
\end{equation}
where $\Pi$ is a permutation matrix such that $123\rightarrow 312$, $R$ permutes $123\to 132$ and $\sum_{k=0}^2(q_k+r_k)=1$ ($q_k,r_k$ can be negative). It is easy to see that $\Pi^k$ and $\Pi^kR$ cover all $6$ permutations of the string $123$. For convenience we put $Q=\sum_{k=0}^2q_k\Pi^k$ and $Q'=\sum_{k=0}^2r_k\Pi^k$ so that
\begin{equation}
S=Q+Q'R\, .    
\end{equation}
Orthogonality mean $SS^T=\mathbb{1}$ and thus
\begin{equation}\label{eq: orthogonality qBvN}
\begin{aligned}
\mathbb{1} &= QQ^T+Q'(Q')^T +QR(Q')^T+Q'RQ^T \\ &= QQ^T+Q'(Q')^T+2QQ'R\, ,
\end{aligned}
\end{equation}
where the last equality is obtained from using the fact that $R\Pi^{-k} = \Pi^{k}R$. This equation can be true if and only if (i) $QQ'=0$ or (ii) $QQ'=R$. However, the latter is not possible because determinants of $Q,Q'$ are +1 whereas the $R$'s determinant is $-1$. We are left with two distinct possibilities: either $Q$ or $Q'$ is zero, giving us $S_+=Q$ or $S_-=Q'R$. The subscript $\pm$ is to indicate the respective solution has determinant $\pm 1$. Note that $S_-$ corresponds to a discontinuous dynamics that is not physical.

In Appendix \ref{appendix: derivation d=3} we show that $S_+$ has a unique form that reads 
\begin{equation}
S_+(\phi) = q_0 I +q_1 \Pi + q_2\Pi^2\, ,
\end{equation}
where $q_k(\phi)=\frac{1}{3}\left(1+2 \text{Re}{\left(\omega^k e^{i\phi}\right)}\right)$, $\omega=e^{i2\pi/3}$ is the cubic root of unity. Note that for any $\phi$ only one of the quasi-probabilities is negative and because $\sum_{k=0}^2\omega^k=0$ we have $\sum_{k=0}^2q_k=1$. As such, this non-deterministic dynamics is reversible as the dynamics it generalizes.  

What we need to fix now is the domain of the state space to which $S_+(\phi)$ is a valid transformation. As $\mathbf{p}$ describes a probability distribution on the lattice it must stay positive for any $S_+(\phi)$. Again, the proof is in the Appendix \ref{appendix: derivation d=3} and we give here the solution:
\begin{equation}
    p_k (\theta) = \frac{1}{3}(1 + t \hat{\mathbf{a}}_k\cdot [\sin\theta, \cos\theta])\, , \quad k=0,1,2\, ,
\end{equation}
where $0\leq t \leq 1$, $0\leq \theta \leq 2\pi$, and $\hat{\mathbf{a}}_k$'s are 2-dimensional real unit vectors such that $\sum_{k=0}^{2} \hat{\mathbf{a}}_k = 0$.  

We arrived at a generalised dynamics for $d=3$, conserving the collision entropy of any initial positive probability distribution $\mathbf{p}$. This is a quasi-bistochastic dynamics that preserves positivity of $\mathbf{p}$ and thus resembles quasi-probability representations of quantum theory discussed in \cite{ferrie2009framed}. Indeed, this dynamics and the set of admissible probability distributions is equivalent to rotations around the axis $\hat{z}$ of qubit states with the Bloch vector $\mathbf{s}=t[\sin\theta,\cos\theta,0]$ confined to the $xy$-plane. 

Dropping the continuity of $S$, we end up with a bigger dynamical system that still describes a qubit but with a larger dynamics that contains experimentally impossible operations. They correspond to reflections of qubit's Bloch vector (including forbidden universal-NOT gate) if we map them to two-dimensional Hilbert space.    

Of course, laboratory measurements on a qubit give only two outcomes (qubit in the state $|0\rangle$ or $|1\rangle$ along the measurement direction), so we need to show how to interpret the lattice probability distribution $\mathbf{p}$. We easily recover the measurement probabilities along an arbitrary direction on the Bloch sphere's equator $\hat{\mathbf{m}}$ if we use the over-completeness of the vectors $\hat{\mathbf{a}}_k$, $\sum_{k=0}^2\hat{\mathbf{a}}_k\hat{\mathbf{a}}_k=\frac{3}{2}I$ (here $\mathbf{a}\mathbf{b}$ denotes a dyadic product of two vectors). We have
\begin{subequations}
\begin{align}
p(\pm|\hat{\mathbf{m}}) &= \frac{1}{2}\left(1 \pm \hat{\mathbf{m}}\cdot\mathbf{s}\right) \\
&= \frac{1}{2}\left(\mathbf{1}\pm \mathbf{v}\right)\cdot\mathbf{p}
\end{align}
\end{subequations}
where $\mathbf{1}$ is vectors of all ones and $\mathbf{v}=[\hat{\mathbf{m}}\cdot\hat{\mathbf{a}}_0,\hat{\mathbf{m}}\cdot\hat{\mathbf{a}}_1,\hat{\mathbf{m}}\cdot\hat{\mathbf{a}}_2]$. We define $\mathbf{e}(\pm|\hat{\mathbf{m}}) = \frac{1}{2}(1\pm \mathbf{v})$ as the {\it effect} corresponding to the measurement along $\hat{\mathbf{m}}$ with outcome $\pm$. Effects formalism is not the primary concern of this paper but there is extensive literature on this topic \cite{short2010strong} the reader can consult. 

In Appendix \ref{appendix: derivation d=4} we show how to extend the continuous dynamics found for $d=3$ to $d=4$. The significance of this extension is a full reconstruction of qubit states and their physical transformations. 

A natural question at this point is if we can derive a two-qubit dynamics and thus, using a set of two-qubit universal gates, dynamics of any $D$-dimensional quantum system. We already know such collision entropy preserving quasi-bistochastic dynamics equivalent to two qubits --- it can be constructed using a SIC-POVM frame \cite{kiktenko2020probability}, a mapping from a four-dimensional Hilbert space to quasi-probabilistic space. This means we can at worst get a larger class of systems, some of which may not correspond to two qubits. What would such systems be? These are for now open questions we will address in the future work.   

\section{Conclusions}

The gist of this paper is that one can deduce the existence of a qubit together with its full dynamics from a deterministic (reversible) dynamics of a particle hopping on a one-dimensional lattice with $4$ vertices if one postulates the collision entropy conservation. Hilbert space is not needed but you can recover it if you need to. 

Using this information-theoretic postulate we get qubit's dynamics as an orthogonal quasi-bistochastic continuous processes of a particle hopping on a one-dimensional 4-vertex lattice. The particle's states are restricted to non-negative probability distributions that can be uniquely mapped to qubit's measurement probabilities.

Our results can be positioned in the ongoing research to derive quantum mechanics from some basic, information-theoretic principles without invoking orthodox Hilbert space axioms. 

Some open questions:
\begin{enumerate}
\item It is not clear at the moment what the physical significance of the collision entropy is. Technically, it enforces orthogonality of the quasi-bistochastic dynamics $S$ and thus its reversibility: $S^{-1}=S^T$. However, you can imagine a more general reversibility where $S^{-1}\neq S^T$. You can also notice that the $S$'s orthogonality is equivalent to the conservation of qubit's purity by unitary dynamics.  
\item How to recover a dissipative qubit dynamics? 
\item Can we get some other, perhaps post-quantum dynamics (for instance, PR-boxes \cite{scarani2006feats}) conserving Renyi entropies for other $\alpha$? 
\item Can we extend this approach to continuous variables systems? 
\end{enumerate}

After finishing this work, we learnt about a paper by Brandenburger et al. \cite{brandenburger2022renyi} where the authors also use Renyi entropies to connect qubit with quasi-probability distributions via quantum uncertainty principle. How Brandenburger et al.'s results are related to ours requires in-depth study.

\section*{Acknowledgements}
P.K. is supported by the Polish National Science Centre (NCN) under the Maestro Grant no. DEC-2019/34/A/ST2/00081. D.K. is supported by the National Research Foundation, Singapore, and A*Star under the CQT Bridging Grant.

\begin{appendix}

\section{Derivation of dynamics and states for $d=3$}
\label{appendix: derivation d=3}

Here, we show the parameterization of the orthogonal quasi-bistochastic matrix for $d=3$. There are two forms of the solution. The first with $+1$ determinant has the form
\begin{equation}\label{eq: 3d +1 qbistochastic matrix form}
S_+ = \begin{bmatrix}
q_0 & q_2 & q_1 \\ q_1 & q_0 & q_2 \\ q_2 & q_1 & q_0
\end{bmatrix}
\end{equation}
with the constraints
\begin{subequations}\label{eq: +1 det quasibistochastic constraint}
\begin{align}
q_0 + q_1 + q_2 = 1\, , \\
q_0^2 + q_1^2 + q_2^2 = 1\, ,\\
q_0q_1 + q_0q_2 + q_1q_2 = 0\, .
\end{align}
\end{subequations}
Parameterizing the solution space based on the constraints above means that we are solving the problem of intersection between 3-dimensional hypersphere with a hyperplane of the same dimension. The general procedure for solving this problem is presented in Appendix \ref{appendix: hypersphere intersection solution}. The solution for the parameter space is
\begin{subequations}\label{eq: 3d quasibistochastic matrix parameters}
\begin{align}
q_0(\phi) & = \frac{1}{3} + \frac{2}{3}\cos \phi\, ,\\
q_1(\phi) & = \frac{1}{3} - \frac{2}{3}\sin(\frac{\pi}{6}+ \phi)\, ,\\
q_2(\phi) & = \frac{1}{3} - \frac{2}{3}\sin(\frac{\pi}{6}- \phi)\, .
\end{align}
\end{subequations}
The above can be compactly expressed as $q_k = q_k(\phi) = \frac{1}{3}\left[1+2\text{Re}(\omega^k e^{i\phi})\right]$, $k=0,1,2$, where $\omega$ is the cube root of unity. These quantities can take values from $-\frac{1}{3}\leq q_k \leq 1$ and clearly satisfies unit sum for all $\phi$. $S_+=S_+(\phi)$ here forms the group of quasi-bistochastic SO(3) matrices.

The other solution is orthogonal quasi-bistochastic matrix with $-1$ determinant of the form
\begin{equation}
    S_{-} = \begin{bmatrix} q_0  & q_1 & q_2 \\ q_1 & q_2 & q_0 \\ q_2 & q_0 & q_1
    \end{bmatrix}
\end{equation}
with similar constraint as in Eq. (\ref{eq: +1 det quasibistochastic constraint}). Consequently, the solution space can be parameterize similarly as in Eq. (\ref{eq: 3d quasibistochastic matrix parameters}). Contrary to the previous case, $S_{-}$ here is not continuous and do not form a group.

The next thing that we will find is the domain of the state space, where we have specifically chose to be nonnegative and will behave consistently under the $S(\phi)$ above. The reason we can do this is due to the self-duality relation between the state and effect \cite{muller2012structure}. This choice of construction will not change the behavior of the system. With this in mind, let us now construct the state space $\mathcal{S}\subset\mathbb{R}^3_+$ based on the quantum dynamic that we just obtained.

The problem of finding the state space can be stated as follows. Suppose that we have a quasi-bistochastic matrix $S_+$ from Eq. (\ref{eq: 3d +1 qbistochastic matrix form}) with $q_k$ given by Eq. (\ref{eq: 3d quasibistochastic matrix parameters}). The goal is then to find the domain $\mathcal{S}\subset \mathbb{R}^3_+$ where $\mathbf{p}\in\mathcal{S}$ satisfies
\begin{equation}
    S_+\mathbf{p} =\mathbf{p}' \in \mathcal{S} \quad \forall \, \mathbf{p}, \, \forall\, \phi\, .
\end{equation}
Since $\mathbf{p} = [p_0, p_1, p_2]^T $ is a probability distribution, it is then constraint to have unit sum, $\sum_k p_k = 1$. Since the matrix can take negative values, the state space $\mathcal{S}$ is then a subset of the probability simplex.

To solve this, we bring up the fact that $S_+$ is an orthogonal matrix so it leaves the squared-norm of the state vector invariant after the transformation, i.e., $|S_+\mathbf{p}|^2=|\mathbf{p}|^2$. A typical probability vector has squared-norm that is less or equal to 1 but we know that $S_+$ can potentially bring probability vector with squared-norm of 1 into negative probability distribution. Therefore, it implies that there exist an upper bound $K < 1$ for the squared-norm of $\mathbf{p} \in \mathcal{S}$. To find this bound, we consider the parameterized state:
\begin{eqnarray}
    \mathbf{r} &=& \lambda \begin{bmatrix}1 \\ 0 \\ 0
    \end{bmatrix} + (1-\lambda)\begin{bmatrix}\frac{1}{3} \\ \frac{1}{3} \\ \frac{1}{3} \end{bmatrix}\nonumber \\
    &=& \frac{1}{3}\begin{bmatrix} 1+2\lambda \\ 1-\lambda \\ 1-\lambda
    \end{bmatrix}\, ,\quad 0<\lambda< 1\,.
\end{eqnarray}
The squared-norm of $\mathbf{r}$ can be easily calculated to be $(1+2\lambda^2)/3$. The goal here is to find the largest $\lambda$ such that 
\begin{equation}
    S_+\mathbf{r}\geq 0 \quad \forall\, \phi\, .
\end{equation}
The above positivity criteria then implies that 
\begin{subequations}
\begin{align}
    \lambda & \geq  \frac{1}{1-3q_0}\, , \\
    \lambda & \geq  \frac{1}{1-3q_1}\, , \\
    \lambda & \geq  \frac{1}{1-3q_2}\, .
\end{align}
\end{subequations}
Since it needs to be in the range $0 < \lambda < 1$ and works for all $\phi$, one can easily deduce that in the end we have $\lambda \geq \frac{1}{2}$. Therefore, we can infer that the squared-norm of $\mathbf{p}$ takes the range $\frac{1}{3} \leq \sum_k p_k^2 \leq K =\frac{1}{2} $. The extremal states (pure states) of $\mathcal{S}$ are $\mathbf{p}\geq 0$, $\sum_k p_k = 1$, with $|\mathbf{p}|^2 = \frac{1}{2}$.

From here we can parameterize $\mathbf{p}$ as
\begin{equation}
    p_k = \frac{1}{3}(1 + t \hat{\mathbf{a}}_k\cdot [\sin\theta, \cos\theta])\, , \quad k=0,1,2\, ,
\end{equation}
where $0\leq t \leq 1$, $0\leq \theta \leq 2\pi$, and $\hat{\mathbf{a}}_k$'s are 2-dimensional real unit vector and $\sum_k \hat{\mathbf{a}}_k = 0$. If we choose $\hat{\mathbf{a}}_0 = [0, 1]$, then it is natural to have $\hat{\mathbf{a}}_1 = [\frac{\sqrt{3}}{2}, -\frac{1}{2}]$ and $\hat{\mathbf{a}}_2 = [-\frac{\sqrt{3}}{2}, -\frac{1}{2}]$. We can re-parameterise $[t\sin\theta, t\cos\theta] \to [x,y]$ with the condition $x^2 + y^2 \leq 1$. It can also be shown easily that $\mathbf{p}$ above is closed under transformation of $S_-$, i.e., $S_{-}\mathbf{p}\in\mathcal{S}$.

With the hindsight of Hilbert space quantum mechanics, we already know the degrees of freedom $[x,y]$ corresponds to the components of the Bloch sphere in unit circle. In fact, with the choice of $\hat{\mathbf{a}}_k$ above, we recovered the trine quasi-probability representation of a qubit in the $xy$-plane \cite{kiktenko2020probability}.

As for the measurement space, one only needs to find the real vector $\mathbf{m}$ that satisfies
\begin{equation}
    0 \leq \mathbf{m}\cdot \mathbf{p} \leq 1 \quad \forall\, \mathbf{p}\in\mathcal{S}\, .
\end{equation}
The shrinking of the state space from the classical simplex and the deformity of the geometry allows the effect space to be larger than the classical effect space, and hence take on negative values. This trade-off is known as (strong) self-duality \cite{muller2012structure}.

\section{Derivation of dynamics and states for $d=4$}
\label{appendix: derivation d=4}

The construction of the extended theory in $d=4$ can be done in a similar manner as how it is done in $d=3$. However, finding the general form for the quasi-bistochastic SO(4) group using the method above can be quite tedious and complicated. Instead, we will construct it through some basic assumptions about its properties. First we note that the quasi-bistochastic SO(3) matrix is a subgroup of the quasi-bistochastic SO(4) matrices. Hence, there exists 4 elementary rotation matrices
\begin{subequations}
\begin{align}
    R_1 & = \begin{bmatrix}
    1 & 0 & 0 & 0 \\ 0 & q_0 & q_2 & q_1 \\ 0 & q_1 & q_0 & q_2 \\ 0 & q_2 & q_1 & q_0
    \end{bmatrix}\, ,\\
    R_2 & = \begin{bmatrix}
    q_0 & 0 & q_2 & q_1 \\ 0 & 1 & 0 & 0 \\ q_1 & 0 & q_0 & q_2 \\ q_2 & 0 & q_1 & q_0
    \end{bmatrix}\, ,\\
    R_3 & = \begin{bmatrix}
    q_0 & q_2 & 0 & q_1 \\ q_1 & q_0 & 0 & q_2 \\ 0 & 0 & 1 & 0 \\ q_2 & q_1 & 0 & q_0
    \end{bmatrix}\, ,\\
    R_4 & = \begin{bmatrix}
    q_0 & q_2 & q_1 & 0 \\ q_1 & q_0 & q_2 & 0 \\ q_2 & q_1 & q_0 & 0\\ 0 & 0 & 0 & 1
    \end{bmatrix}\, ,
\end{align}
\end{subequations}
with $R_k = R_k(\phi)$ and $q_k = q_k(\phi)$ exactly having the same expression in Eq. (\ref{eq: 3d quasibistochastic matrix parameters}). Then, the general quasi-bistochastic SO(4) matrices $S$ can be written in terms of
\begin{equation}
    S = R_1(\phi_1)R_2(\phi_2)R_3(\phi_3)R_4(\phi_4)\, ,
\end{equation}
where $\phi_k$'s are real parameters that can be found for any quasi-bistochastic SO(4) matrices.

We employ the same method to find the state space and find that extremal states have squared-norm $|\mathbf{p}|^2=\frac{1}{3}$. Hence, the parameterization of the extremal state follows a similar form as the $d=3$ case:
\begin{equation}
    p_k = \frac{1}{4}\left(1 + \hat{\mathbf{b}}_k\cdot[x,y,z]\right)\, ,\quad k=0,1,2,3\, ,
\end{equation}
where we have the constraint $x^2+y^2+z^2\leq 1$ and $\hat{\mathbf{b}}_k$'s are 3-dimensional real unit vector that satisfy $\sum_k \hat{\mathbf{b}}_k = 0$. As a generalization from the $d=3$ case that has vectors of equilateral triangle, we then can have $\hat{\mathbf{b}}_k$'s to be vectors of tetrahedron:
\begin{subequations}
\begin{align}
    \hat{\mathbf{b}}_0 & = [0,0,1]\, ,\\
    \hat{\mathbf{b}}_1 & = \left[\sqrt{\frac{8}{9}}, 0, -\frac{1}{3}\right]\, ,\\
    \hat{\mathbf{b}}_2 & = \left[-\sqrt{\frac{2}{9}}, \sqrt{\frac{2}{3}}, -\frac{1}{3}\right]\, ,\\
    \hat{\mathbf{b}}_3 & = \left[-\sqrt{\frac{2}{9}}, -\sqrt{\frac{2}{3}}, -\frac{1}{3}\right]\, .
\end{align}
\end{subequations}
We have recovered the full Bloch vectors $[x,y,z]$ and hence the most elementary system in discrete quantum system --- the qubit. In fact, this quasiprobability representation corresponds to the frame representation with SIC-POVM frames \cite{kiktenko2020probability}.

Lastly, the effect space is also constructed in the similar manner as the previous section.

\section{Solution to the intersection between $n$-dimensional hypersphere and hyperplane}\label{appendix: hypersphere intersection solution}

Here, we will show how to obtain the solution to the intersection of $n$-dimensional hyperplane and $n$-dimensional hypersphere with unit radius. This problem can be formulated as finding the solution space of $\mathbf{a} = [a_1, a_2, \dots, a_n]\in\mathbb{R}^n$ given that it satisfies two equations:
\begin{subequations}
\label{eq: hypersphere and hyperplane equations}
\begin{align}
    \sum_i a_i &= a_1 + a_2 + \dots + a_n = 1\, ,\label{eq: hyperplane unit sum}\\
    \sum_i a_i^2 &= a_1^2 + a_2^2 + \dots + a_n^2 = 1\, .\label{eq: hypersphere unit rad}
\end{align}
\end{subequations}

Before going on how to obtain the solution space, let us first learn how to parameterize the solution for $n$-dimensional hypersphere with radius $r$:

\begin{equation}\label{eq: hypersphere general}
    b_1^2 + b_2^2 +\dots +b_n^2 = r^2\, .
\end{equation}

The trick for this is to iteratively reduce the problem into a 2-dimensional sphere equation, which we already know how to parameterize. Let $x_1^2 = b_1^2$, $y_1^2 = b_2^2 + b_3^2 + \dots + b_n^2$, $r_1= r$ and we have reduced the Eq. (\ref{eq: hypersphere general}) into equation of a circle:
\begin{equation}
    x_1^2 + y_1^2 = r_1^2\, .
\end{equation}
Therefore, the solution to this can be parameterized as 
\begin{equation}
    x_1 = r_1\cos t_1\, , \quad y_1=r_1\sin t_1\, .
\end{equation}
Then, we can repeat the same thing again for
\begin{equation}
    y_1^2 = b_2^2 + b_3^2 + \dots + b_n^2 = r_1^2\sin^2t_1\, .
\end{equation}
Letting $x^2_2 = b_2^2$, $y_2^2 = b_3^2 + \dots + b_n^2$, $r_2^2 = r_1^2\sin^2t_1$, we obtain another level of parameter
\begin{equation}
    x_2 = r_2\cos t_2\, ,\quad y_2 = r_2\sin t_2\, .
\end{equation}
Doing this $n-1$ times will resolve the parameterization problem.

The following describes the method to solve the intersection problem. Suppose that $U = [\mathbf{u}_1, \mathbf{u}_2, \dots, \mathbf{u}_n]$ is an orthogonal matrix with $\{\mathbf{u}_1, \mathbf{u}_2, \dots, \mathbf{u}_n\}$ forming an orthonormal basis with $\mathbf{u}_n = (1,1,\dots, 1)/\sqrt{n}$. We then can write
\begin{equation}
    \mathbf{a} = U\mathbf{b} = \begin{bmatrix}
    \uparrow & \uparrow & & \uparrow \\ \mathbf{u}_1 & \mathbf{u}_2 & \dots &\mathbf{u}_n \\ \downarrow & \downarrow & & \downarrow
    \end{bmatrix}\begin{bmatrix}
    b_1 \\ b_2 \\ \vdots \\ b_n
    \end{bmatrix}\, .\label{eq: solution form for hypercube/plane problem}
\end{equation}
From the above expression, we have the following two equations
\begin{equation}
\begin{cases}
    \frac{1}{\sqrt{n}} = \mathbf{u}_n^T\mathbf{a} = b_n\, ,\\
    b_1^2 + b_2^2 + \dots +b_{n-1}^2 = \sqrt{1-\frac{1}{n}}\, .
\end{cases}
\end{equation}
The second equation becomes a problem of $(n-1)$-dimensional hypercube with radius $r=1-\frac{1}{n}$. The solution can be parameterized using the technique discussed above. Upon parameterizing $\mathbf{b}$, the form of $\mathbf{a}$ in Eq. (\ref{eq: solution form for hypercube/plane problem}) immediately satisfies Eq. (\ref{eq: hypersphere and hyperplane equations})

\end{appendix}

\bibliography{references}

\end{document}